\documentclass[twocolumn,aps,pre,showpacs,preprintnumbers,floatfix,superscriptaddress]{revtex4-1}

\usepackage{color}
\usepackage{amsmath}
\usepackage{amssymb}
\usepackage{graphicx}
\usepackage{xspace}

\newcommand{\eg}{{e.g.,\/}\xspace}
\newcommand{\ie}{{i.e.,\/}\xspace}

\newcommand{\Eq}[1]{Eq.~(\ref{#1})}

\newcommand{\Fig}[1]{Fig.~\ref{#1}}
\newcommand{\Sec}[1]{Sec.~\ref{#1}}

\begin{document}

\title{Multi-frequency Raman amplifiers}
\author{Ido Barth}
   \affiliation{Racah Institute of Physics, The Hebrew University of Jerusalem, Jerusalem 91904, Israel}
   \email{ido.barth@mail.huji.ac.il}
   \affiliation{Department of Astrophysical Sciences, Princeton University, Princeton, New Jersey 08544, USA}
   
\author{Nathaniel~J. Fisch}
   \affiliation{Department of Astrophysical Sciences, Princeton University, Princeton, New Jersey 08544, USA}
\date{\today}

\begin{abstract}
In its usual implementation, the Raman amplifier features only one pump carrier frequency.
However, pulses with well-separated frequencies can also be Raman amplified while compressed in time.
Amplification with frequency-separated  pumps is shown to  hold even in the highly nonlinear, pump-depletion regime, as derived through a fluid model, and demonstrated via particle-in-cell (PIC) simulations.
The resulting efficiency is similar to single-frequency amplifiers,  but, due to the beat-wave waveform of both the pump lasers and the amplified seed pulses,  these amplifiers feature higher seed intensities with a shorter spike duration.
Advantageously, these amplifiers also suffer less  noise backscattering, because the total fluence is split between the different spectral components. 
\end{abstract}

\maketitle

\section{Introduction}
Backward Raman amplifiers (BRA) provide a promising path to the next generation of  short pulse high-intensity lasers that may circumvent the damage limit of conventional materials.
The main idea is to couple a short seed and a long counter-propagating pump through an electron plasma wave (EPW) in such a way that the pump energy is transferred to the seed that is amplified and compressed via Raman backscattering  \cite{Malkin_PRL_99}.
This mechanism  was extensively studied with respect to various physical effects, including wave breaking \cite{Toroker_PoP_14,Edwards_PRE_15,Farmer_PRE_15}, longitudinal and transverse nonlinearities \cite{Malkin_PRE_14,Barth_PRE_16,Malkin_PRL_16}, precursors \cite{Tsidulko_PRL_02}, group velocity dispersion (GVD) \cite{Toroker_PRL_12}, 
inverse bremsstrahlung \cite{Berger_PoP_04}, Landau damping \cite{Malkin_PRE_09,Balakin_PoP_11},
and premature parametric backscattering of the pump \cite{Malkin_PRL_00}.
The mechanism has also enjoyed experimental implementation \cite{Suckewer03,Suckewer04,Suckewer05}.

In particular, what emerged from these studies of physical effects were techniques, exploiting the laser bandwidth, designed to improve the  operation of Raman plasma amplifiers in different regimes.
The exploiting of the bandwidth, and in particular chirping the frequency,  underlies a number of other studies as well \cite{Hur07,Vieux11,Nuter13,Yang15,Balakin16,Mahdian17,Wu15}.
Chirping the pump laser, together with exploiting a density gradient,  can suppress noise and precursors  \cite{Malkin_PRL_00}. 
It can also overcome relativistic saturation  \cite{Barth_PRE_16}.
Chirping the seed pulse, and exploiting GVD, can accommodate a shorter plasma for the same amplification as for an unchirped seed \cite{Toroker_PRL_12}.
An alternative method to suppress backscattering from noise envisions splitting the pump energy over a few frequencies \cite{Balakin_PoP_03}, where, in order to preserve the amplification efficiency,  the allowed frequency spacing is limited by the spectral width of the single-frequency amplified seed.
Noise suppression by multifrequency pulses, has also been suggested for inertial confinement fusion systems, where the extent of penetration without backscattering depends  on the frequency spacing \cite{Barth_PoP_16,Liu_PoP_17}.
Incoherent pump lasers, with not too small a correlation time, can amplify coherent seeds similarly to coherent pump lasers, but with the advantage of less backscattering due to noise  \cite{Edward_PoP_17}.
Experimental studies demonstrated that, in fact, chirping the pump could compensate detuning due to density gradients thereby facilitating the Raman amplification \cite{RenPOP08,YampolskyPOP08,Yampolsky_PoP_11}.

Here we show that a wave packet comprising two or more well-separated carrier frequencies can be amplified with a similar efficiency as a single-frequency pulse even in the nonlinear regime.
We call this regime multifrequency BRA (MFBRA) to distinguish it from the usual single frequency BRA (SFBRA). 
Importantly, in addition to mitigating the premature backscattering of the pump that is  common to other methods that require bandwidth \cite{Malkin_PRL_00,Balakin_PoP_03,Edward_PoP_17}, 
MFBRA is advantageous  because of its beat-wave waveform.
Here the width of each spike in the beat-wave waveform is smaller than the envelope width, a feature that can be used advantageously.
By a proper preparation of the initial phases that takes into account GVD, the peak intensity of the beat-wave can be engineered to be located at the center of the amplified pulse envelope when leaving the plasma, 
thereby producing an output pulse with the same total fluence,  but with a peak intensity higher than would be possible using SFBRA.

This paper is organized as follows.
In \Sec{sec2} we employ the fluid model to show that double frequency BRA (DFBRA) is possible and to analyze the conditions for such amplifiers.
In \Sec{sec3}, PIC simulations are presented, confirming the effect. 
These simulations are used to optimize the amplification of seeds with two or more carrier frequencies.
We summarize our conclusions in \Sec{sec4}.


\section{Doubly three-wave interaction} \label{sec2}

Consider the wave equations for the Raman-scattered electromagnetic (EM) wave and the electron plasma wave (EPW) within the linearized fluid model for unmagnetized homogeneous plasma \cite{Kruer},
\begin{eqnarray}
  \hat{D}_{\rm em} \mathbf{b} &=&  - \omega_e^2 n_e \mathbf{a} \label{D_b}  \\
  \hat{D}_{\rm epw} n_e       &=&   \frac{c^2}{2}\partial_x^2\left(\mathbf{a}\cdot \mathbf{b}\right) \label{D_ne}
\end{eqnarray}
where,
\begin{eqnarray}
\hat{D}_{\rm em}  &=& \partial_t^2+\omega_e^2-c^2\partial_x^2          \\
\hat{D}_{\rm epw} &=& \partial_t^2+\omega_e^2-3 v_e ^2\partial_x^2
\end{eqnarray}
are differential operators for the EM wave and EPW respectively.
In Eqs.~(\ref{D_b})-(\ref{D_ne}), the total EM vector potential is decomposed into a large and stationary pump wave, $\mathbf{a}$, and a small counterpropagating amplified seed, $\mathbf{b}$.
The electromagnetic vector potentials, $\mathbf{a}$ and $\mathbf{b}$, are in the units of $m_e c^2/e$ and the electron density perturbation, $n_e$, is rescaled by the unperturbed density, $n_0$.
Also, $\omega_e^2=4\pi e^2 n_0/m_e$ is the plasma frequency squared;
$e$ and $m_e$ are the electron charge and mass respectively;
$c$ is the speed of light;
and $v_{e}=\sqrt{T_e/m_e}$ is the electron thermal velocity, with $T_e$ being the electron temperature.
In addition, we neglect the ion motion and, as a result, neglect stimulated Brillouin scattering (SBS), 
notwithstanding that SBS can itself produce a laser compression effect 
\cite{Lehmann13,Weber13,Chiaramello16,Schluck16,Edwards_PoP_16}.
This assumption is justified because, for the parameters of interest, stimulated Raman scattering (SRS) is dominant over SBS \cite{Forslund,Edwards_PoP_16}.

For simplicity, we consider only two carrier frequencies, but the generalization to more than two frequencies is straightforward.
We decompose both the pump and  the seed into two spectral components,
\begin{eqnarray}
    \mathbf{a} &=& \Re \left[a_1 e^{i(\omega_{a_1}t+k_{a_1} x)}+a_2 e^{i(\omega_{a_2}t+k_{a_2} x)}\right]\hat{y} \label{a} \\
    \mathbf{b} &=& \Re \left[b_1 e^{i(\omega_{b_1}t-k_{b_1} x)}+b_2 e^{i(\omega_{b_2}t-k_{b_2} x)}\right]\hat{y}, \label{b}
\end{eqnarray}
where $\hat{y}$ is a unit vector in the transverse direction and $\Re$ denotes the real part.
The pump carrier frequency spacing is defined as
\begin{eqnarray} \label{delta}
    \delta=\omega_{a_2}-\omega_{a_1}.
\end{eqnarray}
The seed carrier frequencies, $\omega_{b_1,b_2}$, are downshifted with respect to the pump carrier frequencies, $\omega_{a_1,a_2}$, according to the Raman resonance conditions,
\begin{eqnarray} \label{resonance_w}
    \omega_{b_1,b_2}=\omega_{a_1,a_2}-\omega_{f_1,f_2},
\end{eqnarray}
where, $\omega_{f_1,f_2}$ are the EPW frequencies that are determined by the dispersion relation,
\begin{eqnarray} \label{dispersion_EPW}
    \omega_{f_1,f_2}^2=\omega_e^2+3v_e^2 k_{f_1,f_2}^2.
\end{eqnarray}

\begin{figure}[tb]
\centering
	\includegraphics[width=0.6\linewidth,trim=0cm 0cm 0cm 0cm,clip]{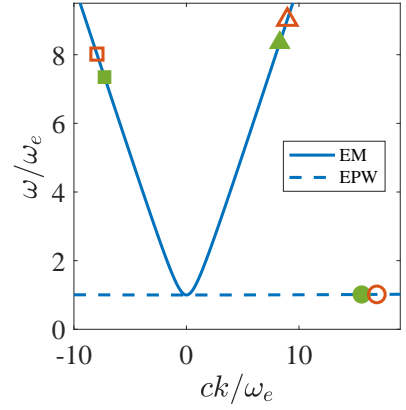}
	\caption{(color online). Illustration of the doubly three-wave interaction in ($\omega-k$) space. 
	The frequencies and wave numbers of the pumps (triangles), the seeds (squares), and the EPW obey the resonance conditions of Eqs. (\ref{resonance_w}) and (\ref{resonance_k}) for both  indices, 1 (filled points) and 2 (empty points).
	The solid and dashed lines are the dispersion curves of the EM wave and the EPW, respectively.}
	\label{fig1}
\end{figure}

Practically, for small $T_e$, we can approximate  $\omega_{f_1,f_2}\approx \omega_e$ and choose the seed frequencies in \Eq{resonance_w}  accordingly, such that their frequency spacing is the same as the pump frequency spacings, \ie $\omega_{b_2}-\omega_{b_1}=\delta$.
The laser wave numbers, $k_{a_1,a_2}$ and $k_{b_1,b_2}$ are determined by the EM dispersion relations,
\begin{eqnarray} \label{dispersion_EM}
    k_{a,b}=\frac{\omega_{a,b}}{c}\sqrt{1-\frac{\omega_e^2}{\omega_{a,b}^2}}.
\end{eqnarray}
The wave number of the EPW is then set by the resonance condition,
\begin{eqnarray} \label{resonance_k}
    k_{f_1,f_2}=k_{a_1,a_2} + k_{b_1,b_2},
\end{eqnarray}
where for our definitions, $k_{a,b}>0$ [see Eqs.(\ref{a})-(\ref{b})].
The underlining assumption here is that the EPW contains two, well separated, carrier frequencies.
\ie the EPW can be decomposed similarly to the EM waves, 
\begin{equation}
   n_e = \Re \left[n_1\, e^{i(\omega_{f_1}t+k_{f_1} x)}+n_2\, e^{i(\omega_{f_2}t+k_{f_2} x)}\right] \label{n},
\end{equation}
where, $n_{1,2}$ are slow complex envelopes.
In \Fig{fig1}, we illustrate an example of two Raman resonance conditions and the dispersion relations in $(\omega - k)$ space, where each set of three waves fulfill both temporal and spatial resonance conditions of \Eq{resonance_w} and \Eq{resonance_k}, respectively.

We continue by substituting Eqs.~(\ref{a}-\ref{b}) into the wave equations (\ref{D_b}-\ref{D_ne}) and use the envelope approximation, \ie neglect the second order derivatives of the wave amplitudes $b_{1,2}$ and $n_{1,2}$.
In the RHS of Eqs.~(\ref{D_b}-\ref{D_ne}) we keep only the resonant terms that obey both temporal [\Eq{resonance_w}] and spatial [\Eq{resonance_k}] resonance conditions.
Besides, we neglect all amplitude derivatives in the RHS of (\ref{D_ne}) since these nonlinear terms are considered small.
After applying the dispersion relations (\ref{dispersion_EPW}) and (\ref{dispersion_EM}) in the left-hand sides (LHS) one gets
\begin{eqnarray}
 &\partial_t b_{1,2} -c_{b_{1,2}} \partial_x b_{1,2} = - \frac{\omega_e^2}{2i\omega_{b_{1,2}}}\, n_{1,2}^{*} \, a_{1,2}  \label{bbb}  \\
& \partial_t n_{1,2} +c_{n_{1,2}} \partial_x n_{1,2} = - \frac{c^2 \left(k_{a_{1,2}}+k_{b_{1,2}} \right)^2}{4i\omega_e}\,a_{1,2}\, b_{1,2}^{*}.
 \label{nnn}
\end{eqnarray}
Here, $c_{b_{1,2}} = c^2 k_{b_{1,2}} /\omega_{b_{1,2}}$ are the the group velocities of the EM waves.
Similarly, $c_{n_{1,2}} = v_e^2 k_{n_{1,2}}/ \omega_e $ are the EPW group velocities, but they are usually negligible in BRA since the EPW is effectively localized relative to the amplified pulse that propagates at nearly the speed of light.
The nonresonant terms that were neglected in the RHS of Eqs.~(\ref{bbb}-\ref{nnn}) contain exponents of the form $\exp[i(k_{a_\alpha}+k_{b_\beta}-k_{n_\gamma}) x]$, where the subindexes, $\{\alpha,\beta,\gamma \}$ are not all the same.
These terms contain fast phases and, therefore, do not contribute to the averaged Raman resonant amplification dynamics.
This assumption holds as long the spacing between the wave numbers of the two EPWs is larger than their width. 

By analysis of the linearized three-wave system [Eqs.~(\ref{bbb})-(\ref{nnn})] one can show that the resonance width of each spectral component is equal to $2\gamma$, 
where, for linear polarization, $\gamma=a_0 \sqrt{\omega_a \omega_e}/2$, is the linear Raman growth rate for pump frequency $\omega_a=\omega_{a_1}$ and initial pump amplitude $a_0$.
Therefore, to avoid overlap between neighboring resonances, the spectral separation condition is 
\begin{eqnarray} \label{condition1}
    \delta > 4\gamma.
\end{eqnarray}
Note that this condition is analogous to the Chirikov criterion for resonance overlap in nonlinear oscillators \cite{Chirikov}.

However, this is not the only condition on the spacing, $\delta$.
Since we consider a short seed, we must guarantee that the beat frequency is large enough such that the seed envelope contains at least one oscillation.
The period of the beat oscillation is  $2\pi/\Omega$, where $\Omega=\delta/2$ is the beat frequency.
Let us define $\tau$ to be the typical duration of the seed, \eg the full width at half maximum (FWHM).
The resulting condition is thus,
\begin{eqnarray} \label{condition2}
	\delta > \frac{4\pi}{\tau}.
\end{eqnarray}

In the next section, we confirm our analysis through PIC simulations that capture the kinetic effects that were neglected in the fluid model and that treat the laser pump and seed  more generally than the envelope approximation utilized above.

\section{PIC simulations} \label{sec3}

\begin{figure}[tb]
	\centering
 	\includegraphics[width=1.0\linewidth,trim=0cm 0cm 0.cm 0cm,clip]{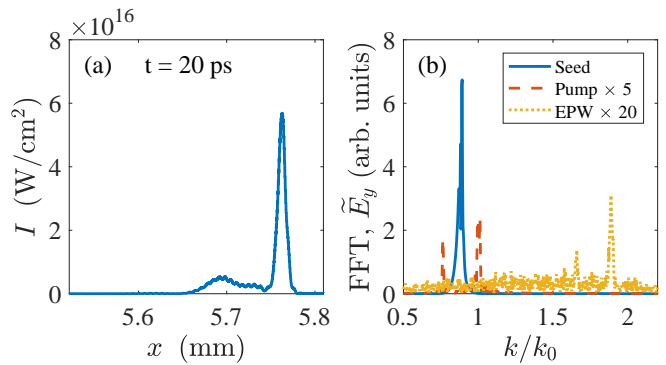}
\caption{(color online). PIC simulation of a SFBRA. The intensity of the amplified pulse (a) is in the nonlinear regime. 
The spectra (b) of the  amplified seed (blue), pump (dashed red) and EPW (dotted yellow) obey the Raman resonance condition,  $k_{f} =k_a+k_b$, where $k_a=k_0$, $k_b=0.89 k_0$, and $k_f=1.89 k_0$.
A secondary Raman backscattering of the seed is also observed  as smaller spikes at $k=0.76 k_0$ in the pump spectrum and $k=1.65 k_0$ in the EPW spectrum.
Some backscattering of the seed  is not unexpected, since the seed reaches an intensity far greater, in fact,  than the pump intensity.} 
	\label{fig2}
\end{figure} 
 
\begin{figure}[t]
	\centering
	\includegraphics[width=1.0\linewidth,trim=0cm 0cm 0.cm 0cm,clip]{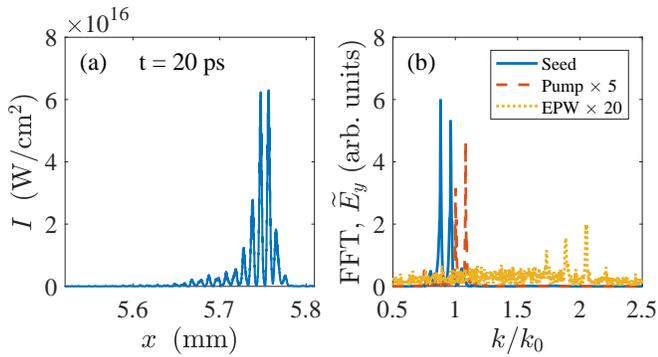}
\caption{(color online). PIC simulation of a DFBRA. 
The intensity of the amplified pulse (a) is in the nonlinear regime with beat-wave structure.
The spectra (b) of the  amplified seed (blue), pump (dashed red) and EPW (dotted yellow) obey the doubly Raman resonance condition, $k_{f_{1,2}} =k_{a_{1,2}}+k_{b_{1,2}}$, where $k_{a_{1}}=k_0$, $k_{a_{2}}=k_0+\delta/c=1.08 k_{0}$, $k_{b_{1,2}}=k_{a_{1,2}}-\omega_e/c=[0.88, 0.96]\times k_0$, and $k_{f_{1,2}}=[1.89, 2.06]\times k_0$.
A secondary Raman backscattering is also observed with smaller spikes.
}
	\label{fig3}
\end{figure}

We employ the PIC code EPOCH \cite{epoch} to  compare  SFBRA (\Fig{fig2}) to DFBRA  (\Fig{fig3}).
We consider a uniform unperturbed electron density of $n_0=2.5\times10^{19} {\rm cm}^{-3}$, electron temperature of $T_e=30$eV, and  immobile ions. 
In the simulations we use $32$ cells per $\mu$m and $16$ particles per cell.
To reduce the simulation time, we employ a window of $0.3$mm in width, moving with the seed. 

In the first example, the (single frequency) pump wavelength is $\lambda_0=800$nm, \ie  $\omega_{0}=2\pi \times 375$ THz and the plasma is underdense with $n_0/n_{\rm cr}=0.0145  $,  where $n_{\rm cr}=1.1 \times 10^{21}/(\lambda_a [\mu m])^2$ is the critical density. 
The pump intensity is $I_0=10^{14} {\rm W/cm}^2$ so the pump dimensionless amplitude, $a_0= 8.5 \times 10^{-10}\lambda_{0}[\mu m] \sqrt{I_0[{\rm W/cm}^2]}$ (for linear polarization) was $a_0=0.068$.
In terms of \Eq{a}, we choose $a_1=a_0$, $a_2=0$, and $\omega_{a_{1}}=\omega_0$.
Due to the resonance condition (\ref{resonance_w}), we downshift the seed frequency by the plasma frequency, $\omega_{b}=\omega_{a}-\omega_{e}= 2 \pi \times 330$ THz, where we neglected the thermal correction.
The seed has Gaussian profile with FWHM of $80$fs. 
In terms of \Eq{b}, we initiate the seed envelopes by the time dependent boundary conditions at  $x=0$ via
\begin{eqnarray}
	b_{1}&=&\bar{b}_{1} e^{-\frac{(t-t_0)^2}{2\sigma^2}}\\
	b_{2}&=&0
\end{eqnarray}
where, $\sigma=34$fs, $t_0=100$fs.
The seed amplitude is the same as the pump amplitude, $\bar{b}_{1}=a_0=0.068$.
As shown in \Fig{fig2}, at time $t=20$ps (\ie when the seed front is at about $5.8$mm inside the plasma) the seed intensity is amplified by a factor of $600$.
The efficiency, in this case, is $\eta=0.75$, \ie 
the pump transferred $75\%$ of its energy to the amplified pulse. 

To illustrate the multifrequency BRA, we introduce, in \Fig{fig3}, a pump that has two frequencies with spacing $\delta$.
Explicitly,  $\omega_{a_1}=\omega_0$ and $\omega_{a_2}=\omega_{0}+\delta$, such that the beat frequency is $\Omega=\delta/2 \ll\omega_0$.
The total pump fluence (energy per cross area) is kept the same as in the single frequency example, but now, it is equally split over the two frequencies,
$I_{a_1}=I_{a_2}=I_0/2=5\times 10^{13}\,{\rm W/cm}^2$ so the dimensionless amplitudes are $a_1=a_2=a_0/\sqrt{2}=0.0048$.
Then, \Eq{a} at the plasma boundary, $x=0$, becomes
\begin{equation}
	\mathbf{a}(x=0,t)= \sqrt{2}\, a_0 \cos\left(\Omega t\right) \, \cos\left(\tilde{\omega}_0 t \right) \hat{y},
\end{equation}
where, $\tilde{\omega}_0=\omega_0+\Omega$ is the fast $(\tilde{\omega}_0 \gg \Omega)$ carrier frequency.
Note that now, for the same pump fluence, the maximum pump intensity is  twice  that of the the previous example.
The seed also comprises two carrier frequencies, $\omega_{b_{1,2}}=\omega_{a_{1,2}}-\omega_{e}$, where, as before, we neglected the thermal correction. 
Therefore, the seed spacing is the same as the pump spacing, \ie $\omega_{b_2}=\omega_{b_1}+\delta$.
For simplicity, we choose both initial envelopes in a Gaussian form, 
\begin{eqnarray}
	b_{1,2} = \bar{b}_{1,2} e^{-\frac{(t-t_0)^2}{2\sigma^2}}e^{i\phi_{1,2}},
\end{eqnarray}
where, $\bar{b}_{1,2}$ are real amplitudes, and $\phi_{1,2}$ are the phases of each spectral component.
To keep the total seed fluence as in the previous example, we choose $\bar{b}_{1,2}=a_0/\sqrt{2}=0.0048$ so the initial seed at the plasma boundary, $x=0$, reads
\begin{eqnarray} \notag
  	\mathbf{b} = \sqrt{2} \, a_0 e^{-\frac{(t-t_0)^2}{2\sigma^2}}\, \cos\left(\tilde{\omega}_b t+  \tilde{\phi} \right)
	\, \cos\left(\Omega t+\varphi \right),
\end{eqnarray}
where, $\tilde{\omega}_b=(\omega_{b_1}+\omega_{b_2})/2=\tilde{\omega}_0-\omega_e$,
$\varphi=(\phi_2-\phi_1)/2$,
$\tilde{\phi}= (\phi_1+\phi_2)/2$,
and, as before, the beat frequency is $\Omega=(\omega_{b_2}-\omega_{b_1})/2=\delta/2$.
In the example shown in \Fig{fig3}, the spacing is $\delta=2\pi \times 30$ THz, which is about $8\%$ of the pump frequency, and no initial phase difference is introduced, \ie $\phi_1=\phi_2=0$. 
All other parameters are kept the same  as in the SFBRA example [\Fig{fig2}].
In this example, the linear growth rate is $\gamma=2\pi \times 3.1$THz, so the separation condition in \Eq{condition1} is met.
Also, the seed FWHS is $\tau=80$\,fs, \ie $4\pi/\tau=2\pi\times 25 \,{\rm THz}<\delta$ as required in \Eq{condition2}.

\Fig{fig3} shows that a seed comprising two carrier frequencies can be Raman amplified if the pump also consists of two frequencies that are upshifted by the plasma frequency.  
This amplification that begins in the linear regime continues in the pump depletion regime despite the nonlinear interaction between the waves.
By connecting the local maxima, we can define the beat-wave envelope.
It is notable that, for the same simulation time, the beat-wave envelope is wider but with a higher peak than that of the single frequency pulse in \Fig{fig2}.
Nevertheless, the efficiency is $\eta=0.64$, which is similar to the efficiency of the SFBRA ($\eta=0.75$).
This difference results from the slower linear stage because of the smaller pump amplitude, $a_{1,2}<a_0$.
However, in the nonlinear (pump depletion) stage, both examples have the same growth rates (slopes).
This means that the rates of the energy transfer from the pump to the seed are equal, \ie the efficiency in the nonlinear stage is the same while the linear stage last different times resulting in a delay in entering the nonlinear stage for DFBRA compare to SFBRA  [see \Fig{fig5}].

The spectra of the amplified seed, the pump, and the EPW, are shown in \Fig{fig3}b.
All of them comprise two dominant frequencies that each triplet fulfills the three-wave interaction resonance condition.
This example demonstrates the mechanism of the two-frequency BRA that was introduced  in \Sec{sec2} and in \Fig{fig1}.
It is clear that, in this example, both seed frequencies, $\omega_{b_{1,2}}$ are independently amplified via a  three-wave interaction of \Eq{resonance_w}.
Importantly, we can conclude that these two resonances remain well separated and the resonance overlap is insignificant also in the nonlinear regime when the linear analysis of \Eq{condition1} is invalid.

\begin{figure}[tb]
	\centering
	\includegraphics[width=1.0\linewidth,trim=0cm 0cm 0.cm 0cm,clip]{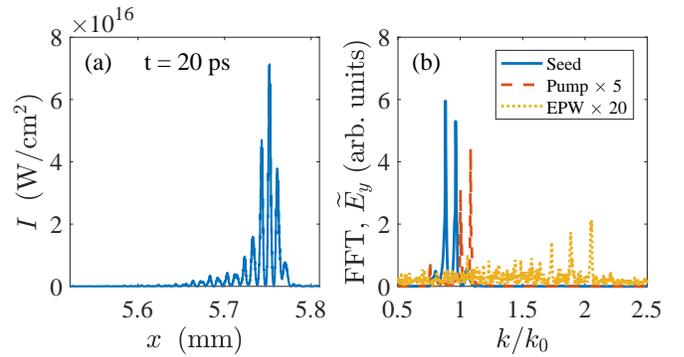}
\caption{(color online). PIC simulation of a DFBRA as in \Fig{fig3} but with an initial phase difference of $\pi$. 
The spectra (b) of the amplified seed (solid blue), pump (dashed red) and EPW (dotted yellow) is similar to that without initial phase difference [\Fig{fig3}] but the intensity of the amplified pulse (a) has a maximum at the center of the beat-wave envelope.}
	\label{fig4}
\end{figure}

To optimize the peak intensity of the amplified pulse, one can manipulate the phases of the seed pulse such that one of the local maxima of the beat-wave would coincide with the maximum of the beat-wave envelope just when it exits the plasma.
However, this is not the case in the example shown in \Fig{fig2}, where the two highest local maxima are located at the shoulders of the beat-wave envelope and are about $20$ percent lower than the envelope maximum. 
Fortunately, such manipulation can be accomplished by taking advantage of GVD \cite{Toroker_PRL_12} that differentiates between the seed spectral components, \ie $c_{b_1}\ne c_{b_2}$  [see \Eq{bbb}].
As a result, the relative phase between the two frequencies changes during the passage of the amplified pulse in the plasma leading to a migration of the location of the highest local maximum inside the beat-wave envelope.
This relative phase can be neutralized by an initial phase difference between the two seed spectral components.
In \Fig{fig4}, we present an example of such manipulation
in which we consider an initial phase difference of $\phi_2-\phi_1=\pi$ between the seed frequencies.
Although it is not the most optimum phase difference, the central peak, at $t=20$ps, is located almost at the maximum of the beat-wave envelope resulting in an increase of about $20\%$ in the maximum intensity.
Notably, both the spectrum and the efficiency are almost the same as the previous example [\Fig{fig3}] where the initial phase difference was zero.

\begin{figure}[t]
	\centering
 	\includegraphics[width=1.0\linewidth,trim=0cm 0cm 0cm 0cm,clip]{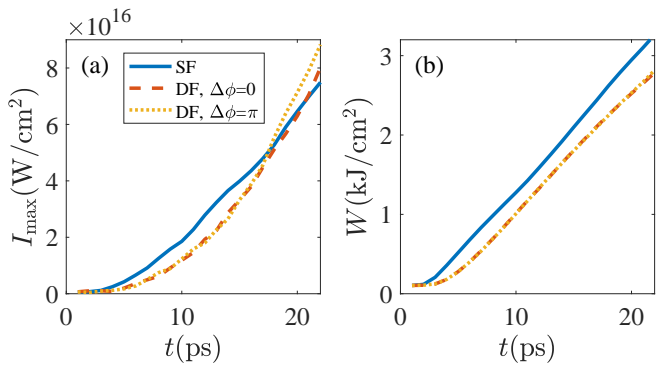}
\caption{(color online) A comparison between the dynamics of SFBRA (solid blue) and of DFBRA with initial relative phase differences of zero (dashed red) and $\pi$ (dotted yellow). Presented are the maximum intensities (a) and total fluences (b) versus the propagation time of the amplified pulse in the plasma.  }
	\label{fig5}
\end{figure}

In \Fig{fig5}, we study the dynamics of SFBRA and DFBRA by comparing the maximum intensities (\Fig{fig5}a) and the total fluences (\Fig{fig5}b) of the amplified pulses in both cases.
The dynamics of the three examples of Figs.~\ref{fig2}--\ref{fig4} are presented versus the propagation time of the amplified pulse.
Notably, the linear growth rate of the SFBRA (blue) is higher than the DFBRA (red and yellow), but the efficiencies in the nonlinear (pump depletion) regime are almost the same.
This can be seen in  \Fig{fig5}b where the  total fluence of the SFBRA grows faster than that of the DFBRA at $t<5$ps, but, at later times, both systems have similar slopes of fluence versus time.

It is notable that although the maximum intensity of the SFBRA is higher than that of the DFBRA at smaller times, the beat-wave waveform of the DFBRA has a higher peak intensity at later stages, $t>17$ps.
Moreover, as shown in \Fig{fig4}, by optimization of the phase between the two carrier frequencies, the peak intensity can be even higher at the time when the amplified pulse exits the plasma (\eg $t=20$ps).
It is clear that the efficiency does not depend on the phase difference between the two seed frequencies but the locations of the local maxima in the envelope change in time due to GVD.
GVD also causes the superimposed oscillations of the  monotonically increasing maximum intensity.
Notably, the difference between the two examples is a result of the phase difference between the seed frequencies, which is zero in the first case and $\pi$ in the second one.
Usefully, one can design the initial seed phases according to the plasma length and density, which determine the phase accumulation of the spectral components. 
It particular, it can be arranged that at the plasma edge a local maximum would coincide with the global maximum of the beat-wave envelope.
\begin{figure}[tb] 
	\centering
 	\includegraphics[width=1.0\linewidth,trim=0cm 0cm 0cm 0cm,clip]{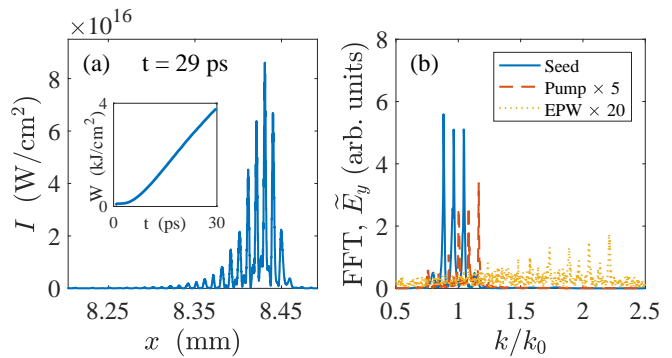}
\caption{(color online) PIC simulation of a MFBRA comprising three carrier frequencies.
The intensity of the amplified pulse (a) is in the nonlinear regime with beat-wave form of three frequencies. 
The total fluences versus the propagation time is plotted in the inset of panel (a).
The spectra (b) of the  amplified seed (blue), pump (dashed red) and EPW (dotted yellow) obey the Raman resonance condition,  $k_{f_{1,2,3}} =k_{a_{1,2,3}}+k_{b_{1,2,3}}$, where, $k_{a_{1,2,3}}=[1,1.08,1.16]\times k_0$, 
$k_{b_{1,2,3}}=[0.88, 0.96,1.04]\times k_0$, and $k_{f_{1,2,3}}=[1.88, 2.04,2.2]\times k_0$.
A secondary Raman backscattering is also observed with smaller spikes.
}
	\label{fig6}
\end{figure}

Finally, we note that pulses with more than two frequencies can also be Raman amplified in a similar way.
In this case, more care should be taken to avoid resonance overlap when many frequencies are involved, and we leave it to future work.
Nevertheless, we present, in \Fig{fig6}, an example of seed and pump that comprises three, evenly spaced, carrier frequencies, $\omega_{a_3}=\omega_{a_2}+\delta=\omega_{a_1}+2\delta$.
Similarly, for the seed frequencies, $\omega_{b_3}=\omega_{b_2}+\delta=\omega_{b_1}+2\delta$, where each pair fulfill the Raman three-wave resonance, $\omega_{b_i}=\omega_{a_i}-\omega_e$ for $i=1,2,3$.
In this example we used the same laser and plasma parameters as in the previous examples, \eg  $\omega_{a_1}=\omega_0$, $\delta=2\pi \times 30$ THz.
However, to maintain the same total fluence, we used a smaller intensity per spectral component,
$I_{a_{1,2,3}}=I_0/3 = 3.33 \times 10^{13}\,{\rm W/cm}^2$.
Additionally, in analogy to \Eq{condition2}, we choose a longer seed, $\tau_{\rm FWHM}=120$fs (instead of $80$fs previously) that can contain the triple-frequency beat-wave waveform of the seed. 


\section{Conclusions} \label{sec4}
In summary, we show that a multifrequency seed can be amplified  and compressed by using a multifrequency pump with the same frequency spacing.
In the linear regime, a simple fluid model suggests that multifrequency BRA is possible, provided that sufficiently large spacing is employed, \ie $\delta>\gamma$, where $\gamma$ is the growth rate.
Moreover, PIC simulations show that the multifrequency amplification continues in the nonlinear regime with similar efficiencies as the SFBRA.
We could not predict this fact from the linearized fluid model.

The advantages of amplifying multifrequency pulses are as follows.
First, similar to other spectral approaches, such pulses experience a reduced premature reflectivity of the pump due to a smaller linear growth rate of each spectral component.
However, uniquely to MFBRA, the secondary backscattering of the amplified seed  is also reduced since the seed  also comprises multiple carrier frequencies.
Therefore, the total (unwanted) reflectivity is reduced, and the amplification efficiency increases.
Second, the duration of each spike in the beat-wave envelope is smaller than that of the envelope, and thus, one can get a much shorter pulse without additional compression. 
Third, by engineering the initial phases of the seed components, the maximum intensity can be optimized for the same efficiency.
Additionally, following the recent study that found  that the total critical intensity for self-focusing might be higher for multicolor beams \cite{Sukhinin_PRA_17}, 
we expect to find a similar delay in the transverse filamentation instability since the same nonlinear Kerr term is responsible for both effects.
Such a delay might enable longer amplification before encountering this transverse instability,  making MFBRA even more favorable over SFBRA.

Although a relatively large bandwidth  is required in MFBRA due to the spacing conditions in Eq.~(\ref{condition1}) and Eq.~(\ref{condition2}), 
there are considerable benefits in having a shorter spike in the beat-wave form, a higher peak intensity,   noise suppression, and the possibly reduced transverse instability.
We also  note that similar multifrequency amplification might also be realized for Brillouin amplifiers, but further work is required to verify the separation conditions between possible resonances.
Also, although we consider here linear polarizations, similar results are predicted for circularly polarized waves or linearly polarized waves, but with perpendicular polarization.
These types of waves have the property of reduced parasitic backscattering \cite{Barth_PoP_16}, but since they do not have a beat-wave waveform,  no improvement in the maximum intensity is expected.
These results should also carry over to using a multifrequency plasma seed instead of a seed laser \cite{Qu_PRL_17} or to pulses with nonzero orbital angular momentum \cite{Arteaga_PoP_2017}.

\acknowledgments{
This work was supported by
NNSA Grant No.~DE-NA0002948 and
AFOSR Grant No.~FA9550-15-1-0391.
}



\begin{thebibliography}{99}

\bibitem{Malkin_PRL_99}
    V.~M. Malkin, G. Shvets, and N.~J. Fisch, Phys. Rev. Lett. {\bf 82}, 4448 (1999).
    
\bibitem{Toroker_PoP_14} Z. Toroker, V.~M. Malkin, and N.~J. Fisch, Phys. Plasmas {\bf 21}, 113110 (2014).

\bibitem{Edwards_PRE_15} M.~R. Edwards, Z. Toroker, J.~M. Mikhailova, and N.~J. Fisch, Phys.Plasmas {\bf 22}, 074501 (2015).

\bibitem{Farmer_PRE_15} J.~P. Farmer and A. Pukhov, Phys. Rev. E {\bf 92}, 063109 (2015).    
    
\bibitem{Malkin_PRE_14} V.~M. Malkin, Z. Toroker, and N.~J. Fisch, Phys. Rev. E {\bf 90}, 063110 (2014).

\bibitem{Barth_PRE_16} I. Barth, Z. Toroker, A.~A. Balakin, and N.~J. Fisch, Phys. Rev. E {\bf 93}, 063210 (2016).

\bibitem{Malkin_PRL_16} V.~M. Malkin and N.~J. Fisch, Phys. Rev. Lett. {\bf 117}, 133901 (2016).

\bibitem{Tsidulko_PRL_02} Y.~A. Tsidulko, V.~M. Malkin, and N.~J. Fisch, Phys. Rev. Lett. {bf 88}, 235004 (2002).

\bibitem{Toroker_PRL_12} Z. Toroker, V.~M. Malkin, and N.~J. Fisch, Phys. Rev. Lett. {\bf 109}, 085003 (2012).

\bibitem{Berger_PoP_04} R. L. Berger, D. S. Clark, A. A. Solodov, E. J. Valeo, and N. J. Fisch, Phys. Plasmas {\bf 11}, 1931 (2004).

\bibitem{Malkin_PRE_09} V.~M. Malkin and N.~J. Fisch, Phys. Rev. E {\bf 80}, 046409 (2009).

\bibitem{Balakin_PoP_11} A.~A. Balakin, N.~J. Fisch, G.~M. Fraiman, V.~M. Malkin, and Z. Toroker, Phys. Plasmas {\bf 18}, 102311 (2011).

\bibitem{Malkin_PRL_00}V.~M. Malkin, G. Shvets, and N.~J. Fisch, Phys. Rev. Lett. {\bf 84}, 1208 (2000).

\bibitem{Suckewer03} Y.~Ping, I.~Geltner, A.~Morozov,  N.~J.~Fisch,  and S.~Suckewer, Phys. Rev. E, {\bf 66}, 046401 (2002).
           
\bibitem{Suckewer04} Y.~Ping, W.~Cheng, S.~Suckewer, D.~S.~Clark, and N.~J.~Fisch, Phys. Rev. Lett. {\bf 92}, 175007 (2004).

\bibitem{Suckewer05} W.~Cheng, Y.~Avitzour, Y.~Ping, S.~Suckewer, N.~J.~Fisch, M.~S.~Hur, and J.~S.~Wurtele, Phys. Rev. Lett. {\bf 94}, 045003 (2005).

\bibitem{Hur07} M.~S. Hur, D.~N. Gupta, and H. Suk, J. Physics D-Applied Physics  {\bf 40}, 5155 (2007).

\bibitem{Balakin16} A. A. Balakin, I. Y. Dodin, G. M. Fraiman, and N. J. Fisch, Phys. Plasmas  {\bf 23}, 083115   (2016).

\bibitem{Mahdian17} Z. Mahdian,  S. Mirzanejhad, T. Mohsenpour, and M. Taghipour, OPTIK   {\bf 149}  372 (2017).

\bibitem{Yang15} X. Yang, G. Vieux, E. Brunetti, et al., Scientific Reports {\bf 5} 13333 (2015).  

\bibitem{Vieux11} G. Vieux, A. Lyachev, X. Yang, et al., New J. Phys. {\bf 13}, 063042 (2011).

\bibitem{Nuter13} R. Nuter and V. Tikhonchuk, Phys. Rev. E {\bf 87} 043109 (2013).

\bibitem{Wu15} Z. H. Wu,  X. F. Wei, Y. L. Zuo, et al., Chinese Physics B {\bf 24} 014211 (2015).

\bibitem{Balakin_PoP_03} A.~A. Balakin, G.~M. Fraiman, N.~J. Fisch, and V.~M. Malkin, Phys. Plasmas {\bf 10}, 4856 (2003).

\bibitem{Barth_PoP_16} I. Barth and N.~J. Fisch, Phys. Plasmas {\bf 23}, 102106  (2016).

\bibitem{Liu_PoP_17}  Z.~J. Liu, C. Y.~Zheng, L.~H. Cao, B. Li, J. Xiang, and L. Hao, Phys. Plasmas {\bf 24}, 032701 (2017).

\bibitem{Edward_PoP_17} M.~R. Edwards, K. Qu, J.~M. Mikhailova, and N.~J. Fisch,  Phys. Plasmas {\bf 24}, 103110 (2017). 

\bibitem{RenPOP08} J. Ren, S. Li, A. Morozov, S. Suckewer, N. Yampolsky, V.~M, Malkin, and N.~J. Fisch, Phys. Plasmas {\bf 15}, 056702 (2008).

\bibitem{YampolskyPOP08} N.~A. Yampolsky, N.~J. Fisch, V.~M. Malkin, E.~J. Valeo, R. Lindberg, J.~S. Wurtele, J. Ren, S.~Li, A. Morozov, and S. Suckewer, Phys. Plasmas {\bf 15}, 113104 (2008).

\bibitem{Yampolsky_PoP_11} N.~A. Yampolsky and N.~J. Fisch, Phys. Plasmas {\bf 18}, 056711 (2011).

\bibitem{Kruer} W.~L. Kruer, {\it The Physics of Laser Plasma Interactions} (Addison-Wesley, Reading, MA, 1988).

\bibitem{Lehmann13} G. Lehmann and K. Spatschek, Phys. Plasmas {\bf 20}, 073112 (2013).

\bibitem{Weber13} S. Weber, C. Riconda, L. Lancia, J.-R. Marque`s, G.~A. Mourou, and J. Fuchs, Phys. Rev. Lett. {\bf 111}, 055004 (2013).

\bibitem{Chiaramello16} M. Chiaramello, F. Amiranoff, C. Riconda, and S. Weber, Phys. Rev. Lett. {\bf 117}, 235003 (2016).

\bibitem{Schluck16} F. Schluck, G. Lehmann, C. Muller, and K. Spatschek, Phys. Plasmas {\bf 23}, 083105 (2016).

\bibitem{Edwards_PoP_16} M.~R. Edwards, Q. Jia, J.~M. Mikhailova, and N.~J. Fisch, Phys. Plasmas {\bf 23}, 083122 (2016).

\bibitem{Forslund} D.~W. Forslund, J.~M. Kindel, and E.~L. Lindman, Phys. Fluids {\bf 18}, 1002 (1975).

\bibitem{Chirikov} B.~V. Chirikov, Phys. Rep. {\bf 52}, 263  (1979).

\bibitem{epoch} T.~D. Arber, K. Bennett, C.~S. Brady, A. Lawrence-Douglas, M.~G. Ramsay, N.~J. Sircombe, P. Gillies, R.~G. Evans, H. Schmitz, A.~R. Bell, and C.~P. Ridgers, Plasma Phys. Controlled Fusion {\bf 57}, 113001 (2015).

\bibitem{Sukhinin_PRA_17} A. Sukhinin, A.~B. Aceves, J.~C. Diels, and L. Arissian, Phys. Rev. A {\bf95}, 031801(R) (2017).

\bibitem{Qu_PRL_17} K. Qu, I. Barth, and N.~J. Fisch, Phys. Rev. Lett. {\bf 118}, 164801 (2017).

\bibitem{Arteaga_PoP_2017} J.~A. Arteaga, A. Serbeto, J.~T. Mendonca, K. H. Tsui, L.~F. Monteiro, Phys. Plasmas {\bf 24}, 123108 (2017).

\end{thebibliography}
\end{document}